\documentclass{ws-p8-50x6-00}

\begin{document}

\title{Baryonic Dark Matter: Limits from HST and ISO}

\author{Gerard Gilmore}

\address{Institute of Astronomy, Madingley Rd, Cambridge CB3 0HA, UK
\\E-mail: gil@ast.cam.ac.uk}

\maketitle

\abstracts{Recent HST and ISO observations provide very severe limits
on any compact baryonic contributions to galactic (dark) halos. When
combined with Milky Way Galaxy microlensing results, almost the entire
plausible range of massive compact baryonic objects is excluded by direct
observation.  Deep direct imaging at 7$\mu$m and 15$\mu$m with ISOCAM
on the ISO spacecraft directly exclude hydrogen-burning stars of any
mass above the hydrogen-burning limit, and of any chemical abundance, from
being the predominant explanation of the dark halos of external spiral
galaxies. In the Milky Way Galaxy, HST has provided luminosity
functions to the hydrogen-burning limit in several globular
clusters. The resulting mass functions do not provide any support for
dominance by very low-mass stars. This is consistent with field
surveys for sub-stellar mass brown dwarfs, which show such objects to
be relatively rare. These results are complemented by very deep HST
luminosity functions in the Large Magellanic Cloud, providing strong
support for the (near)-universality of the stellar mass function.
Very recent HST results are available for the nearby dSph galaxy
UMi. This galaxy, the most dark-matter dominated object known on kpc
scales, has a normal stellar mass function at low masses.  The
prospects are bright for dark elementary particles.}

\section{Introduction}

Low mass stars and sub-stellar mass brown dwarfs provide the only
objects which are known to exist in potentially interesting numbers
which might make up a substantial part of the dark matter required in
galactic halos by dynamical arguments. \footnote{Although very old
white dwarf stellar remnants have been considered, contraints from the
combination of an artificially contrived initial mass function, and
over-production of all of luminosity, helium, and later supernovae by
their progenitors, exclude them as astrophysically plausible. }

The number of stars as a function of mass, corrected for stellar
evolutionary and chemical abundance effects -- the Initial Mass
Function (IMF) -- is one of the two key functions defining the
evolution of galaxies and the distribution of (baryonic) matter. (The
star formation history is the other.) There is no {\em ab
initio} understanding of either of these two functions. One therefore
may speculate on arbitrarily complex variations, but ultimately must
be guided by observations. Recent observations indicate an unexpected
but now well-established similarity in the IMF over the full range of
astrophysical sites in which it can be determined directly or
indirectly.

The IMF of high-mass stars -- those which would be the progenitors of
any putative population of 0.5M$_{\odot}$ very old white dwarf stellar
remnants -- has been determined recently, from HST observations in
Local Group galaxies and HIPPARCOS observations in the Solar Neighbourhood.
In spite of the wide range of local chemical abundances and star
formation rates, the Local Group (Massey 1998) and Solar Neighbourhood
(Brown 1998) IMFs are not distinguishable. In more extreme
environments only indirect analyses are possible. Nonetheless, even in
extreme starburst galaxies (Leitherer 1998) the derived high-mass IMF
is standard. Thus, the high-mass IMF does not seem to be a (strong)
function of the local star formation rate. Since chemical element
creation yields from supernovae are progenitor-mass dependent, one may also
determine the IMF of the first stars to form in the Milky Way from
the element ratio pattern in surviving old low-mass stars. Here again,
the evidence (Wyse 1998) is consistent with a standard IMF. Thus, the
high-mass IMF does not seem to be a (strong) function of the chemical
abundance of the gas from which it forms, over some 2-3 orders of
magnitude.

Given the short lifetimes of high-mass stars, their IMF can be
relevant to the dark matter problem only indirectly, through their
low-mass remnants. Most of the mass in the observed standard IMF is
locked up in long-lived low-mass stars, all of which ever formed still
exist as stars today. Thus, the  low mass IMF can be determined
directly, for stars above the minimum mass for hydrogen burning. For
even lower mass stars, which rapidly cool and fade to extremely low
luminosity, recent surveys have also provided direct determinations.
But only locally, given the intrinsic low luminosities.
Nonetheless, recent near and mid infra-red surveys (Lada, Lada, \& Meunch
1998) in regions of recent or continuing star formation determine the
low mass stellar IMF to be adequately described by the standard KTG field
IMF (Kroupa, Tout \& Gilmore 1993). The KTG IMF, which continues to be
uncertain, predominately due to very small number statistics (Reid
1998), but possibly due to real small-amplitude variability (Scalo
1998), is a representation of the local field stars: thus it represents
a time-averaged IMF over the history of the local disk, for solar
abundance stars. The implication is that the low mass IMF has not
shown (strong) variations at high abundances over cosmic time, in
agreement with the conclusions for high-mass stars above.
Importantly, the KTG mass function shows a flattening at low masses,
below about 0.2M$_{\odot}$, such that the total mass is dominated by
hydrogen-burning stars. This unimportance of brown dwarfs for the mass
budget seems well established at high abundances, being supported by
the ongoing field brown dwarf identification programs from the DENIS
large area IR survey (Tinney (1998) private communication).

All these constraints leave open one window of opportunity for compact
baryonic dark matter: low masses, low chemical abundances and old
ages. Primordial chemical abundance, {\em ie} H+He, is the expected
value if the as-yet unidentified baryons required to exist by standard
Big Bang nucleosynthesis arguments form compact objects at early
times. This requirement is almost unavoidable, for compact objects,
since the dark matter seems to be found exclusively in galactic halos,
and not galactic disks (eg Gilmore 1989). Thus, it must have been
locked into a non-dissipative form early. \footnote{An obvious
explanation, that the missing baryons remain distributed in a
(possibly chemically-enriched) inter-galactic medium, is a natural
though as-yet unproven explanation for the fate of the baryons not
found in galaxies and clusters today if they exist in a non-compact
form.}

We now consider, in turn, observational limits on the number of
compact, low-mass, low-abundance baryonic objects in the Milky Way
halo, in other Local Group galaxies, and in more distant galaxies.

\section{The IMF in Galactic Globular Clusters}

HST has revolutionised determinations of the relative numbers of stars
as a function of lumnosity - the Luminosity Function (LF) -- in
Galactic globular clusters.  While conversion of an observed luminosity
function into an IMF remains very model-dependent (d'Antona 1998),
differential comparisons of observed luminosity functions between
clusters of not dis-similar chemical abundances is a very robust check
on possible variations. 

Cool (1998) has reviewed the extensive recent results, justifying the
general conclusion that, considering the range of chemical abundances,
internal cluster structural parameters, and differing Galactic orbits,
any differences between cluster luminosity functions are certainly not
large. In particular, in a beautiful experiment by Piotto, Cool \&
King (1997), four globular clusters with similar chemical abundances
were studied with HST. Three had indistinguishable luminosity
functions, the fourth a substantial {\em deficit} of low mass
stars. The deficit cluster is on an orbit close to the Galactic disk,
providing an explanation for the lost low-mass stars as being tidally
removed from the parent, following the normal internal dynamical
process of mass segregation inside the cluster. Under this
interpretation, all the clusters had a universal luminosity -- and
hence mass -- function, corresponding to the present day luminosity
function of the three similar clusters.

It is worth emphasising here that the only detected difference between
the clusters corresponds to even fewer low mass stars in one case --
ie, even less relevance to baryonic dark matter. Nonetheless, if the
IMF can vary down, it could vary up. It is also possible, at least in
principle, that the three similar clusters are all stripped of low
mass stars to the same extent, and the true IMF is indeed dominated by
now lost low-mass stars. The next step is therefore to
confirm the tidal stripping argument for the variable luminosity
function. Elson etal (1999) determined HST luminosity functions for
several other clusters, and combined them with extant published
data. The results are shown in Figure~1 here.

\begin{figure}[t]
\epsfxsize=25pc 
\epsfbox{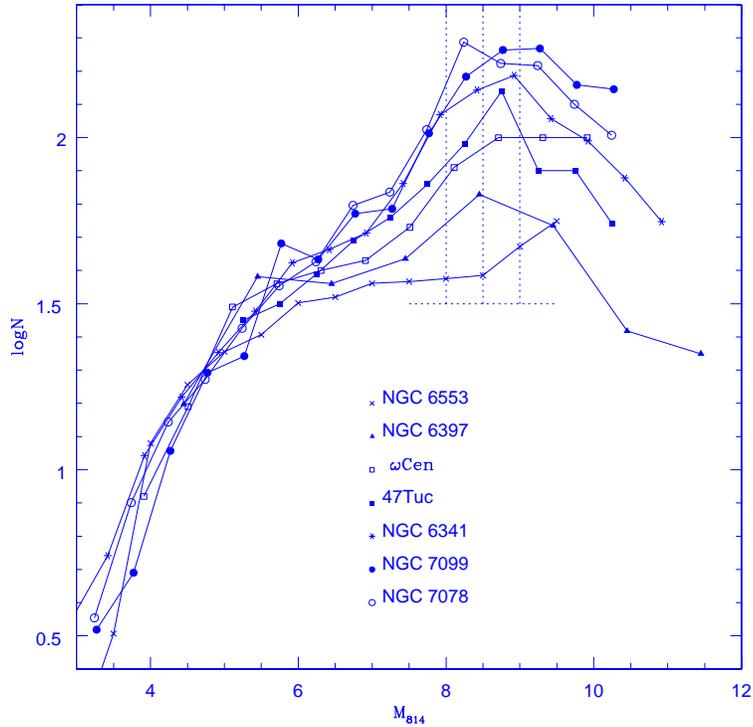} 
\caption{Comparison of luminosity functions for  a set of globular
clusters from HST data, taken from Elson etal (1999). Clear
differences are evident, suspected to correspond to differences in
tidal mass loss and mass segregation histories, and not to different
initial mass functions.
\label{fig:f-1}}
\end{figure}

\begin{figure}[t]
\epsfxsize=20pc 
\epsfbox{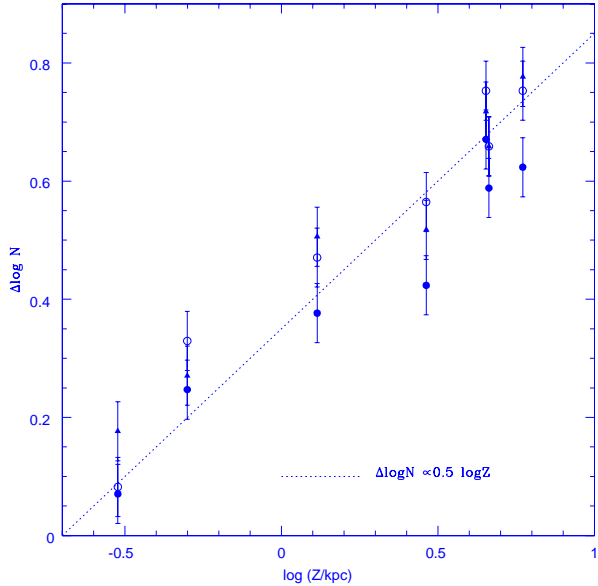} 
\caption{Residuals from the standard line of Figure~1 are shown, as a
function of the logarithmic distance of each cluster from the Galactic
Plane. This latter quantity is a robust measure of the importnace of
the Galactic tidal field. The correlation evident implies that all
observed variations in cluster luminosity functions are artefacts
of varying tidal mass loss and mass segregation histories. There is no
evidence for any variability in the underlying stellar IMF.
\label{fig:f-2}}
\end{figure}

Clear differences between the cluster luminosity functions are
evident. To quantify these, and determine any dependance on the
Galactic tidal field, the distance of each observed function from the
dotted line in Figure~1 was measured. Figure~2 illustrates the
clear dependance of this deficit on logarithmic distance from the Galactic
plane, which is a measure of the tidal field. The relationship is
remarkably good, given that the present position of a cluster is only
a single estimate of its full orbital and tidal history. The resulting
logarithmic distance dependance is convincing evidence that internal
mass segregation and tidal stripping generate the observed differences
between cluster luminosity functions. No case is seen in which an {\em
excess} of low-mass stars is apparent in the data.

That is, to at least first
order, the metal poor globular clusters provide evidence for an
invariant IMF, which has too few low mass stars to be of relevance to
the dark matter problem.

\section{The IMF in Globular Clusters in other galaxies}

The work reviewed above provides clear evidence for an approximately
universal IMF at low masses in old galactic globular clusters. Might
the IMF change with time? with chemical abundance? with parent galaxy
type? To answer these (and related) questions, a large HST study
(project 7307, PI: G. Gilmore) is underway, to provide information
comparable to that available for Galactic globular clusters on 8
globular clusters in the LMC. These clusters span a much wider range
in age and chemical abundance than is available in the Galaxy.

\begin{figure}[t]
\epsfxsize=20pc 
\epsfbox{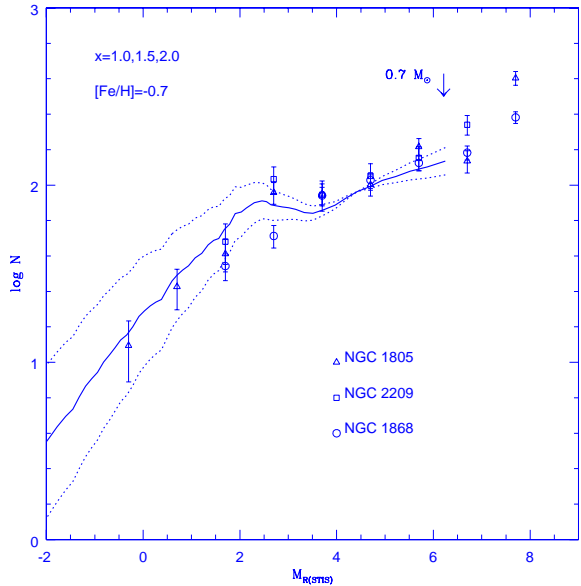} 
\caption{Preliminary luminosity functions for three LMC globular
clusters with a wide age range, from HST data, taken from Elson etal (1999). 
Mass functions with power-law slopes (x=1,1.5,2), converted to expected
luminosity functions, are overplotted. The x=1.5 curve is a good
description of stars in the Solar Neighbourhood. Any systematic differences in
the IMF between the clusters in the LMC and the local stars is small.
\label{fig:f-3}}
\end{figure}

A foretaste of the many results of this study is shown in Figure~3,
which presents luminosity functions for 3 LMC clusters, covering a
very wide age range. Overplotted are expected luminosity functions for
three different IMF slopes. The better fitting slope corresponds to
that of the KTG solar neighbourhood function over the relevant mass
range.

Even from these preliminary data, it is apparent that the luminosity
functions are, to first order, similar to each other and to the
luminosity function of the Milky Way Galaxy. That is, there is no
clear evidence to support any variability in the IMF in globular
clusters as a function of chemical abundance (over a range of 2dex),
as a function of time (over a range of 12Gyr), or as a function of
location (from old halos to modern disks).

\section{The IMF of field stars in other galaxies}

There is no dark matter in globular clusters. Even so,
the common distance and very similar chemical abundances
of all their stars makes them the most suitable
places where accurate determinations of stellar luminosity functions
can be made. What one really would like of course is to know the
stellar luminosity function in the field halo, where dark matter is
predominant. There the implied mass:light ratios are so high that
extreme luminosity functions are required to provide baryonic dark
matter. 

\begin{figure}[t]
\epsfxsize=25pc 
\epsfbox{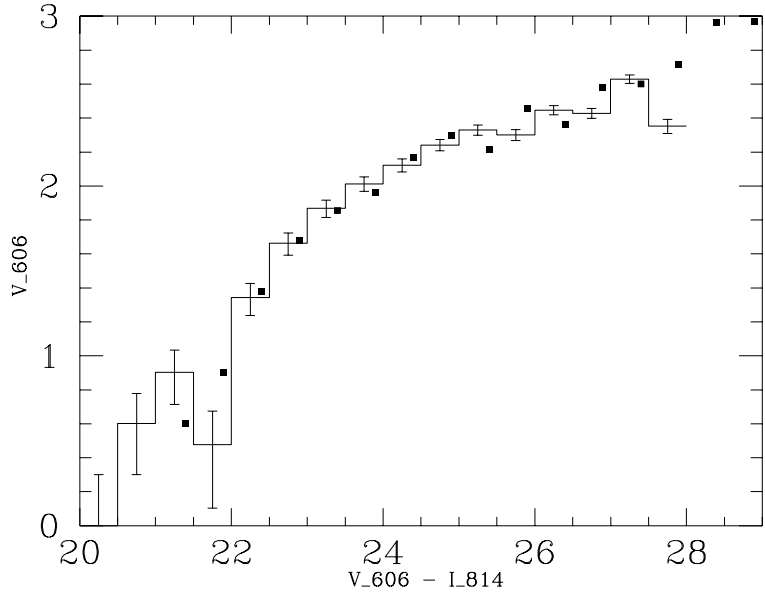} 
\caption{Preliminary luminosity function for the dSph dark matter
dominated galaxy UMi, from Feltzing, Gilmore \& Wyse (1999). The
galaxy data (histogram) is overplotted with data (points) for the
Galactic globular cluster M92, which has similar age and chemical
abundances, and contains no dark matter. The similarity of the two
functions is evident, providing very strong evidence that the stellar
mass function is not related to the total mass:light ratio of its
environment.
\label{fig:f-4}}
\end{figure}

In the Galactic halo, direct HST searches by many authors all agree
on the relevant star count data (see, eg, Gilmore (1997) for a review).
The interpretation is also (fairly) common, that direct HST data
exclude low mass hydrogen-burning stars as candidates for the dark
halo of the Milky Way. Substellar mass objects (brown dwarfs) are
excluded by the microlensing data. Given the difficulty in
understanding the microlensing data, however, one would like some
independent confirmation.

The ideal case is provided by the dSph galaxy UMi. This Milky Way
satellite galaxy is sufficiently far from the Sun that all its stars
are at essentially the same distance, but close enough for one to see
to low masses directly. Its stellar population is approximately a
single age and has a narrow chemical abundance range, with both being
similar to that of the Galactic globular cluster M92.

Most importantly, UMi has an extremely high mass:light ratio of about
60 (Hargreaves, Gilmore, Irwin \& Carter 1994). Thus, any explanation
of dark matter involving low-mass stars implies a  very steep
luminosity function in UMi, compared to M92. This comparison has been
made in HST project 7419 (PI: R.F.G. Wyse). Preliminary results, based
on a sub-part of the data only, are shown in Figure~4, taken from
Feltzing, Gilmore \& Wyse (1999). It is evident that the luminosity
functions of UMi -- perhaps the lowest stellar density system known,
and yet with the highest dark-matter dominance -- 
and M92 -- among the most dense stellar systems, with no dark matter,
are not very different. This provides direct evidence that the stellar
luminosity function does not vary between two environments which
differ in mass:light ratio by a factor of 30 or so. That is, the stars
do not dominate the mass in UMi.

\section{The Dark Halos of Spiral Galaxies: ISO observations}

\begin{figure}[t]
\epsfxsize=25pc 
\epsfbox{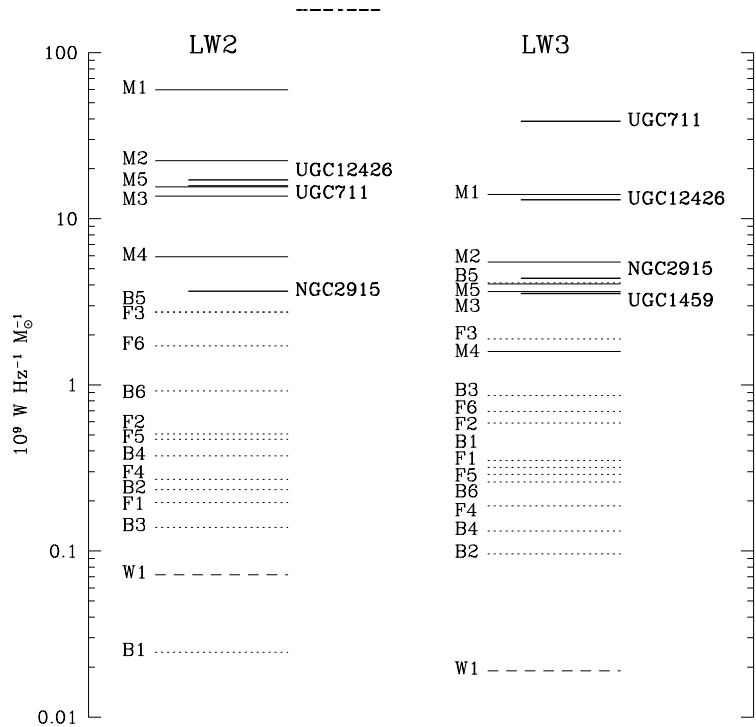} 
\caption{Limits on the stellar populations potentially generating the
extended dark halos of external spiral galaxies, derived from direct
imaging at 7$\mu$m (left column, LW2) and 15$\mu$m (right column, LW3)
by ISOCAM, taken from Gilmore \& Unavane (1999). The solid lines with
adjacent galaxy identifications indicate the observational upper
limits. The various models are on the left of each column. Those with
a solid bar correspond to various mass functions for very low-mass but
hydrogen-burning stars; those with dotted limits are for various age
and mass function brown dwarf models.  Stars at or above the
hydrogen-burning limit, and even young brown dwarfs, are directly
excluded from dominance of the dark matter halos of spiral galaxies.
\label{fig:f-5}}
\end{figure}

The strongest direct evidence for local dark-matter dominance is in the
outer parts of spiral galaxies, where extended gas rotation curves
imply a very extended mass distribution. The various HST studies of
field stars, and the gravitational microlensing surveys, study a
pencil-beam through this halo in the Milky Way. These observations can
be complemented by direct observations of the halos of external
galaxies with known rotation curves. In this case, any observation
samples a column through the entire dark halo. It is well known that
such studies see no extended luminosity, hence the name
`dark'. However, low mass stars are of very low optical luminosity, so
that the direct optical limits are not a strong limit on the mass in 
such a population.

Low mass hydrogen burning stars are however very red, and relatively
luminous in the mid-infrared. The Infrared Space Observatory provided
the ideal mid-infrared imaging system, ISOCAM. Gilmore \& Unavane
(1999) utilised this to observe four edge-on spiral galaxies, and,
after considerable data reduction and some modelling described in that
reference, were able to derive strict limits excluding hydrogen
burning stars of any mass from dominance of the mass in the dark halos
of these galaxies. Their results are summarised in Figure~5.

\section{Summary}

Recent HST observations show remarkable evidence that the stellar
initial mass function is surprisingly universal. The IMF shows no
dependence on chemical abundance over a factor of 100; no dependence
on formation epoch, over 12Gyr; no dependence on type or luminosity of
parent galaxy; no dependence on local stellar density; no dependence
on the total local mass content. To an excellent approximation, any
old stellar population has always a mass:light ratio of a few, 2-3 in
usual visual light solar units. The observed mass:light ratios of a
few tens and higher seen in dwarf galaxies, and outer spiral halos,
are not generated by a varying number of low mass stars.

\begin{figure}[t]
\epsfxsize=20pc 
\epsfbox{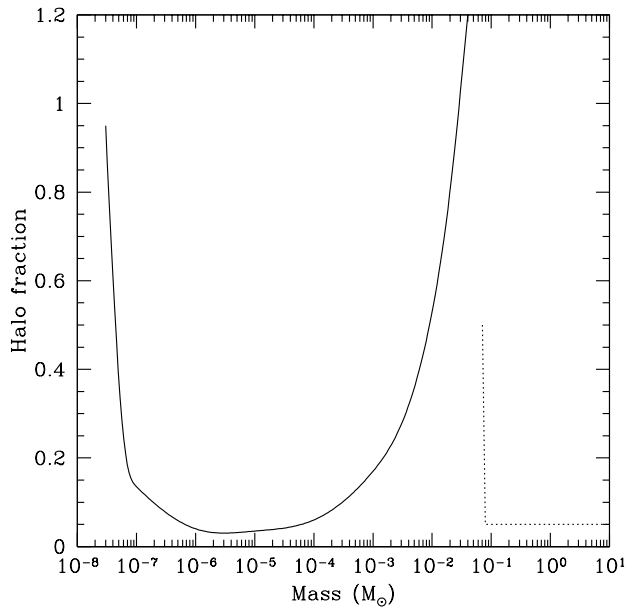} 
\caption{Summary limits on the mass of baryonic compact objects which
might have contributed to the extended dark halos of spiral
galaxies. Masses below the hydrogen burning limit (solid line) are
excluded by the gravitational microlensing surveys. Higher masses
(dotted line) are excluded by HST and ISO observations.  Note that
very old white dwarfs are not excluded directly in the ISO 
analysis, but are excluded by other astrophysical arguments.
\label{fig:f-6}}
\end{figure}

The Gilmore \& Unavane ISO results effectively exclude
hydrogen-burning stars of any mass from generating the dark matter
implied by spiral galaxy rotation curves. Their limits may be combined
with those of the microlensing groups studying dark matter in our own
Galaxy, which exclude very low mass objects. The resulting limits are
combined in Figure~6. Clearly, the mass range which is populated by
normal compact baryonic objects does not provide the dark matter in
normal galaxies.

\section{Conclusion}

Low mass stars do not provide a substantial part of the answer to the
dark matter problem. A large fraction of the total Universal baryon
content which is derived from Big Bang nucleosynthesis models is
apparently not inside galaxies, or in the identified intra-cluster
medium. It is plausibly waiting discovery in the warm Inter-Galactic
Medium (IGM).  The dark matter content of the Milky Way, and its
satellite galaxies, is apparently some other dark stuff.  The future
is bright for particle physicists.


\begin{thebibliography}{99}
\bibitem{aa}d'Antona, F., in {\it The Stellar Initial Mass Function} eds
G. Gilmore \& D. Howell, p157 (ASP Conf Ser vol 142: ASP, Provo) 1998.

\bibitem{ab}Brown, A., in {\it The Stellar Initial Mass Function} eds
G. Gilmore \& D. Howell, p45 (ASP Conf Ser vol 142: ASP, Provo) 1998.

\bibitem{ac}Cool, A., in {\it The Stellar Initial Mass Function} eds
G. Gilmore \& D. Howell, p139 (ASP Conf Ser vol 142: ASP, Provo) 1998.

\bibitem{eg}Elson, R.A., Tanvir, N., Gilmore, G., Johnson, R.A. \&
Beaulieu, S. in {\it Stellar Populations in the Magellanic Clouds},
Victoria, Canada, August 1998 in press 1999.

\bibitem{fgw}Feltzing, S., Gilmore, G., \& Wyse, R.F.G. in preparation 1999.


\bibitem{gg1}Gilmore, G., in {\it Identification of Dark Matter IDM1}
ed N.J.C. Spooner  p73, (World Scientific, Singapore) 1997

\bibitem{gg}Gilmore, G., in {\it Baryonic Dark Matter}, eds
D. Lynden-Bell \& G. Gilmore, p137 (Kluwer, Dordrecht) 1989.


\bibitem{gu}Gilmore, G., \& Unavane, M., \Journal{MNRAS}{in}{press}{1999}.

\bibitem{hg}Hargreaves, J., Gilmore, G., Irwin, M., \& Carter, D.,
\Journal{MNRAS}{271}{693}{1994}.

\bibitem{ktg}Kroupa, P., Tout, C., \& Gilmore,
G. \Journal{MNRAS}{262}{545}{1993}.

\bibitem{al}Lada, E.A., Lada, C.J., \& Muench, A, in {\it The Stellar
Initial Mass Function} eds 
G. Gilmore \& D. Howell, p107 (ASP Conf Ser vol 142: ASP, Provo) 1998.

\bibitem{al1}Leitherer, C., in {\it The Stellar Initial Mass Function} eds
G. Gilmore \& D. Howell, p61 (ASP Conf Ser vol 142: ASP, Provo) 1998.

\bibitem{am}Massey. P., in {\it The Stellar Initial Mass Function} eds
G. Gilmore \& D. Howell, p17 (ASP Conf Ser vol 142: ASP, Provo) 1998.

\bibitem{pck}Piotto, G., Cool, A.M. \& King,
I.R. \Journal{AJ}{113}{1345}{1997} 

\bibitem{ar}Reid, I. N., in {\it The Stellar Initial Mass Function} eds
G. Gilmore \& D. Howell, p121 (ASP Conf Ser vol 142: ASP, Provo) 1998.

\bibitem{as}Scalo, J., in {\it The Stellar Initial Mass Function} eds
G. Gilmore \& D. Howell, p201 (ASP Conf Ser vol 142: ASP, Provo) 1998.

\bibitem{aw}Wyse, R.F.G., in {\it The Stellar Initial Mass Function} eds
G. Gilmore \& D. Howell, p89 (ASP Conf Ser vol 142: ASP, Provo) 1998.


\end{thebibliography}
\end{document}